# A general framework for conformal transformations in electron optics

G. Ruffato[1,*], E. Rotunno[2], V. Grillo[2,**]

[1]University of Padova, Department of Physics and Astronomy 'G. Galilei', via Marzolo 8, 35125 Padova, Italy

[2]CNR-Istituto Nanoscienze, Centro S3, Via G Campi 213/a, I-41125 Modena, Italy

*gianluca.ruffato@unipd.it

**vincenzo.grillo@unimore.it

## Abstract

The implementation of *log-pol* transformation has recently introduced a new boost in electron optics with charged matter vortices, allowing to map conformally between linear and orbital angular momentum (OAM) states and to measure them. That coordinate change belongs to the general framework of Hossack's transformations and it has been recently realized efficiently by means of electrostatic elements. In this letter we show that it is a general property of those conformal transformations to be produced by harmonic phase elements and therefore to admit an electrostatic implementation in the electron optics scenario. We consider a new kind of conformal mapping, the circular-sector transformation, which has been recently introduced for OAM multiplication and division in optics, and discuss how it represents a general solution of Laplace's equation, showing the analogy of the generating phase elements with projected multipole fields and linear charge distributions. Moreover, we demonstrate its capability to perform the sorting of multipole wavefronts, discovering a novel and effective method to measure the strength and orientation of a dipole field in a fast, compact and direct way.

**Manuscript**

The development of structured matter waves is among the most intriguing frontiers of electron optics and electron microscopy so far [1]. It has opened a new horizon of experiments and possibilities in electron microscopy including magnetic characterization of material [2][3], controlled diffraction, aberration and interferometry [4][5][6][7]. The revolution has been started with the introduction of electron vortex beams [8][9][10] carrying an azimuthal phase term $\exp(i\ell\vartheta)$, which suggested the orbital angular momentum (OAM) (denoted by the quantum number $\ell$) as a new electron microscopy quantity. However, the most spectacular and ambitious use of this concept is the OAM sorter that allows for the measurement of this quantity [11][12].

As all structured wave ideas, the OAM sorter is transposed from light optics [13][14] but assumes a different functional meaning in electron microscopy. The somewhat surprising effect of the sorter is to produce, in the stationary phase approximation, a conformal transformation between Cartesian and *log-polar* coordinates. This means that a difficult measurement like the OAM characterization transforms into a simple linear momentum analysis that is typically carried out using a lens.

This conformal transformation is obtained by a first phase element, the so-called *un-wrapper*, imparting a phase contribution that upon propagation alters the amplitude profile, unwrapping the input annular distribution into a linear one. A second element, the so-called *phase-corrector*, removes the phase that was added in the new coordinate framework [13]. Conversely, if the path is analyzed in reverse, the latter element performs the inverse optical transformation (*wrapper*) while the former corrects for the phase-distortions introduced during propagation [15]. As in the case of the optical counterpart [16][17][18], the OAM sorter for electrons has been recently realized by means of holograms [11] introducing the prescribed transformation and correction phases.

However, it is somehow surprising that such complicated phases patterns can be created by a simple set of electrodes that have been indeed put in place experimentally [19][20].

The phase produced by a limited set of electrodes is proportional to the electrostatic potential $V$ integrated along the propagation direction and cannot be arbitrary, but is limited by the constraint that $\nabla^2 \phi = 0$, namely that the function is harmonic, except for the electrodes areas. In passing we notice that a set of magnetic elements can also produce the same phase since in general

$$\phi = C_E \int V dz + \frac{e}{\hbar} \int A_z dz \tag{1}$$

Again, the condition $\nabla^2 \phi = 0$ is still valid. However, this is a very limited subset of phase landscapes that does not even cover elementary effects like defocusing or spherical aberration. It is worth noting that this condition is realized for both OAM sorter elements, i.e. transformer (unwrapper) and phase-corrector (wrapper) [10]. At the same time the concept of conformal mapping is so revolutionary that we may search for a generalization of this idea in order to measure more relevant quantities.

Therefore, this paper aims to study a broader area of conformal mappings, their relation to harmonic phase plates and the kind of quantities they permit to measure in electron microscopy.

The general framework for coordinate transformation has been provided, once again, in optics. In the paraxial regime, the electron wave function $\psi_z(u,v)$ after a propagation length $z$ from an optical element with phase function $\Omega(x,y)$, is given by the following formulation of the Fresnel-Kirchhoff integral [21]:

$$\psi_z(u,v) = \frac{e^{ikz}}{i\lambda z} \int_{-\infty}^{+\infty}\int_{-\infty}^{+\infty} \psi_0(x,y) e^{i\Omega(x,y)} e^{-ik\frac{xu+yv}{z}} dxdy \qquad (2)$$

$\psi_0(x,y)$ being the wavefunction distribution illuminating the phase element, $k = 2\pi/\lambda$ the wavevector modulus, $(x,y)$ and $(u,v)$ the Cartesian reference frames located at the input and output planes, respectively. According to the stationary phase approximation [22], a two-dimensional integral with the form in eq. (2) can be approximated by the values of the argument around the saddle points $\{(x^*, y^*)\}$ of the total phase function $\Phi$, as it follows:

$$\psi_z(u,v) = \sum_{\{(x^*,y^*)\}} \psi_0(x^*, y^*) e^{i\Phi(x^*,y^*)} \frac{2\pi i\sigma}{\sqrt{|AB-C^2|}} \qquad (3)$$

where $A = \partial^2\Phi/\partial x^2$, $B = \partial^2\Phi/\partial y^2$, $C = \partial^2\Phi/\partial x\partial y$, and $\sigma = \text{sgn}(A)$ when $AB > C^2$, $\sigma = -1$ otherwise. In the specific case, the stationary condition on the phase function in eq. (2) leads to the following relation between the gradient of the phase $\Omega(x,y)$ and the point $\vec{\rho} = (u(x,y), v(x,y))$:

$$\vec{\rho} = \frac{z}{k}\nabla\Omega \qquad (4)$$

Therefore, the phase element performs a mapping between a point $(x,y)$ on the input plane, and a point $(u(x,y), v(x,y))$ on the destination plane. Eq. (4) implies that $\vec{\rho}$ is irrotational, i.e. $\nabla \times \vec{\rho} = 0$ [23]. As a consequence, the following condition is satisfied:

$$\frac{\partial u}{\partial y} = \frac{\partial v}{\partial x} \tag{5}$$

A particularly interesting case is provided when the mapping is described by a complex transformation $\rho_c = u(x, y) + iv(x, y)$ which is conformal. Then, in the complex function theory, eq. (5) can be fulfilled by an anti-holomorphic function satisfying the following identity on the Wirtinger operator:

$$\frac{\partial \rho_c}{\partial \bar{\zeta}} = \frac{1}{2}\left(\frac{\partial}{\partial x} + i\frac{\partial}{\partial y}\right)\rho_c = 0 \tag{6}$$

where $\bar{\zeta} = x - iy$. Thus, in addition to eq. (5), the real and imaginary parts $(u, v)$ of the anti-holomorphic complex function $\rho_c$ satisfy the following Cauchy-Riemann condition:

$$\frac{\partial u}{\partial x} = -\frac{\partial v}{\partial y} \tag{7}$$

Due to the seminal work published by Hossack and coworkers in 1987 [22], we will refer to this kind of conformal (anti-holomorphic) mappings as Hossack's transformations. A conformal mapping must conserve the local orthogonality of coordinates direction. As a consequence of previous Cauchy-Riemann equations, the Jacobian matrix $J_\rho$ is symmetric, i.e. $J_\rho^T = J_\rho$, and it can be demonstrated that $J_\rho^T J_\rho = -\det(J_\rho)^{-1} I$, $I$ being the identity matrix and $\det(J_\rho) = -(\partial u / \partial x)^2 - (\partial u / \partial y)^2$. Then, the integral in eq. (2) is indeed a unitary transformation in the quantum mechanics sense.

After combining eq. (4) and eq. (7), the following condition on the phase function $\Omega(x, y)$ is obtained:

$$\nabla^2 \Omega = 0 \qquad (8)$$

Therefore, the phase function of an optical element performing a conformal anti-holomorphic transformation must be harmonic, i.e. a solution of Laplace's equation.

The general statement at the core of this paper is that every Hossack's transformation is essentially based on harmonic phase elements and it inherently exhibits an electrostatic counterpart. As a matter of fact, since $\nabla^2 \phi = 0$ everywhere except on the electrodes, then the desired transformation can be reproduced by means of a finite combination of electrodes (or currents).

It is here worth mentioning the main Hossack's transformations with an application in optics to date beyond the aforementioned *log-polar* case. The first case is a spiral version of the *log-polar* transformation used for improved vortex spectroscopy (but only for radially smooth beams) [24]. The second is a family of polar mappings that perform circular-sector transformations (Fig.1), the combination of which permits to multiply or divide the OAM quantum number of vortex beams [25][26]. The latter case, in particular, assumes a broader relevance in the context of the present study. As a matter of fact, a solution of Laplace's equation (eq. (8)) in 2D for the phase function $\Omega$ in polar coordinates $(r, \vartheta)$ is given by:

$$\Omega(r, \vartheta) = A \cdot r^m \cdot \cos(m\vartheta + \vartheta_0) \qquad (9)$$

where $m$ is an integer. The relation provided by eq. (4) allows obtaining the conformal transformation achieved by the corresponding optical element, i.e. the functional relation between the two polar reference frames $(r,\vartheta)$ and $(\rho,\varphi)$ on the input and output planes, respectively:

$$(\rho,\varphi) = \left(A \cdot m \cdot r^{m-1}, (1-m)\vartheta - \vartheta_0\right) \tag{10}$$

With the substitution $m = 1 - 1/n$, eq. (9) provides exactly the functional dependence of a phase element performing a conformal circular-sector transformation by a factor $n$, as described by eq. (10), that is $\varphi = \vartheta/n$ (plus an angular shift $\vartheta_0$) (see Supplementary Information S1 for detailed calculations). The corresponding phase-corrector is the element performing the inverse transformation, i.e. mapping a circular sector with angular amplitude $2\pi/n$ onto the whole circle (Fig. 1). Its phase function is provided by a circular-sector transformer under the substitution $n \to 1/n$, then it is described by a phase profile as in eq. (9) with multipole order $m' = 1 - n = m/(m-1)$. The trivial circular-sector transformation, given by $n=+1$, is provided by the axially symmetric phase element $\Omega(r,\vartheta) = A \cdot \log(r/b)$. Again, this is a solution of eq. (8), and it completes the set of general solutions for 2D Laplace's equation in polar coordinates provided by eq. (9).

These considerations lead to two important claims. Since the general solution of Laplace's equation in polar coordinates is represented by a circular-sector transformation, we can conclude that any conformal transformation with cylindrical symmetry can be represented by the linear combination of circular-sector transformations. Secondly, the analogy with Laplace's equation in the electromagnetic framework allows us to conclude that the same transformations can be implemented for charged matter waves, e.g. electron beams, by means of proper and spatially limited distributions of charges or currents.

In Fig. 1, the phase patterns of *n*-fold circular-sector transformations are shown for different values of the parameter *n*, applied to a vortex wavefunction with OAM $\ell=+2$. For positive *n*, it is clear the analogy between the phase-correctors and the projected potential generated by electric multipoles. As a matter of fact, for $m<0$ it can be demonstrated (see Supplementary File S2) that the phase in eq. (9) is proportional to the integral in *z* of the potential $V(\bar{s}) \propto As^{m-1} P_{|m|}(\cos\gamma)\cos(m\vartheta)$ generated by an in-plane electric $2^{|m|}$-pole, being $\bar{s}=(\bar{r},z)$, $\gamma$ the angle formed by $\bar{s}$ with the *z*-axis, $P_{|m|}(\cdot)$ the *|m|*-th order Legendre polynomial. This result provides a straightforward method to generate the same phase patterns with an electrostatic techniques, using the corresponding in-plane electric multipole, as shown in Fig. 2(a.1) and (b.1) for a dipole and a quadrupole, implementing the phase-correctors of a +2-fold and a +3-fold circular-sector transformation, respectively.

On the other hand, the phase patterns of the corresponding transformers exhibit a discontinuity along the negative *x*-axis, i.e. at $\vartheta=\pi$. Again, by applying the electro-optical analogy and solving the corresponding Poisson's equation, it can be demonstrated (see Supplementary Information S3) that the phase-pattern resembles the projected electrostatic potential generated by a linear projected charge density distributed along $\vartheta=\pi$:

$$\rho_L(r) = Q_L \cdot r^{-1/n} \cdot \delta(\vartheta - \pi) \tag{11}$$

where $Q_L = -2\varepsilon_0 Am\sin(m\pi)$. It is worth noting that the generated field can be described as the one generated by a fractional-order multipole as defined in [27][28]. For practical reasons, the charge distribution will be limited to a maximum radius value $R$, defining the extent of the electrostatic element. This truncation introduces a discontinuity in the radial direction, imposing the need for an additional ring charge distribution, as shown in Fig. 2. The projected charge density varies on the ring as $\rho_R(\vartheta) = Q_R \cos(m\vartheta)\delta(R-r)$, where $Q_R = 2\varepsilon_0 A(m+1)R^{m-1}$ [33] (Supplementary Information S3). The whole electrode can be designed properly in order to ensure the charge neutrality condition that allows for the convergence of the integral [29].

Analogously, for negative values of $n$, the phase-correctors resemble the integrated field of magnetic multipoles, and the corresponding projected potentials (Fig. 2(c.1) and (d.1)) can be generated by only a ring distribution with the proper $m$ integer value.

In Fig. 2, for each transformation with integer $n$ in the range {-3, -2, +2, +3}, we have reported for the transformer the appropriate electrode configuration made of a linear charge distribution and a ring one. The corresponding projected potential is shown to correctly reproduce the desired phase pattern, as expected.

To summarize, the sources can be essentially distinguished in three categories, depending on the value and sign of $n$ ($m$): 1) internal electrodes generating an in-plane multipole configuration (positive fractional $n$, $n=+1$ included, i.e. integer $m \leq 0$), 2) ring charge distribution or external electrodes (negative fractional $n$, $n=-1$ included, i.e. integer $m \geq 2$), 3) ring charge distribution coupled to a linear one (integer $n$, i.e. fractional $0 < m < 2$).

In practical realizations, controlling the charge distribution can be quite complicated because experimentally it is much more common to control the electrodes shapes and therefore the equipotential lines. The inverse problem of calculating such geometry from the aimed charge distribution has been however tackled with different approaches [30][31]. Alternatively, the use of currents as sources of magnetic fields provides a more direct way to control the phase that does not require the solution of an inverse problem [32] .

As stated above, it is also interesting to notice that the family of circular-sector transformations [25][26] is a "complete basis" [34]. Thus, all Hossack's transformations with cylindrical symmetry are expected to be obtained as a combination of circular-sector transformations. This hints on the idea that an appropriate distribution of needles and multipoles in both transformer and corrector planes can be used to generate a vast category of phase elements. This is therefore a strong alternative to the approach proposed by Verbeeck [35] to introduce programmable phase plates in electron microscopy, with the advantage that most of the beam area is free of electrodes. Thus, they can represent the ideal elements for advanced and versatile measurements and/or beam shaping.

For a given optical transformation, it is worth calculating, if there exists, the family of input fields which are mapped into plane waves. As a matter of fact, a Fourier transform applied in sequence allows to separate them and to perform a sorting of the fields entering the transformer. For instance, in the case of the *log-pol* optical transformation, such a family is represented by beams with an azimuthal phase gradient (vortex beams) and the sequence of *log-pol* transformer and Fourier transform has been widely demonstrated to work as an OAM sorter, as mentioned above. In the specific case of the circular-sector transformation, we found the surprising result that this family is represented by fields endowed with projected multipole phase profiles.

In the specific case of electrons, the family of wavefunctions that transforms into plane waves is

$$\psi(r,\vartheta) = \exp(ikz) \cdot \exp\left(iAr^m \cos\left(m(\vartheta - \vartheta_0)\right)\right) \tag{12}$$

As stated above, a simple $z$ integration demonstrates that this is the phase of an electron that has interacted with an in-plane electric (or magnetic) multipole potential. If we consider an input wavefunction as in eq. (12) and we apply the circular sector transformation with a factor $n=-1/m$, according to eq. (3) the following output is obtained:

$$\psi_z(\rho,\varphi) \propto C \cdot \exp(iA'\rho\cos(\varphi + m\vartheta_0)) \tag{13}$$

which is endowed with a linear phase term, with spatial frequency $A' = Ab^m/a$, $(a, b)$ being the scaling parameter of the applied transformation (see Supplementary Information S4 for detailed calculations). The far-field at a distance $d$ will exhibit a spot forming an angle $-m\vartheta_0$ with the $x$-axis positive direction, and shifted of $dA'/k$ with respect to the origin. For instance, a $n=-1/2$ circular-sector transformation followed by a Fourier transform is able to transform an external quadrupole ($m=+2$) phase object into a point, allowing to measure the quadrupole strength $A$ and its orientation $\vartheta_0$, as shown in Fig. 3 (left panel). In practice this would be an extremely handy method to measure the astigmatism by reading its value and orientation directly on the screen. For example, we envisage that this could be combined with a Zemlin scheme based on tilt series for the measurement of all higher order aberrations very quickly [36].

In a more general case, by performing the scan on a sequence of $(-1/m)$-circular-sector transformation, it is possible to perform a multipole analysis of the input field in an effective and compact way. In particular, it would be of great interest to measure the strength of a dipole term, i.e. $m=-1$, by transforming the dipole phase profile into a linear one. This is achieved by exploiting the logarithmic radial transformation, corresponding to the trivial circular-sector transformation defined by $n=+1$. The required phase element corresponds to the projected potential of an electric monopole located at the origin (Fig. 4). In Fig. 3 (right panel), the transformation of an input dipole phase profile into an output spot is shown for different values of dipole moment and in-plane orientation. As expected, the axial displacement of the far-field spot increases with the dipole strength, while its angular position depends on the rotation angle.

This result appears similar to what found for the *log-polar* case where a broad peak in the OAM spectrum could be associated to the dipole strength (but not orientation). However that method showed a strong limitation in the case of low magnetic moment due to the discrete character of OAM. In this case the complete linearity of the displacement with the magnetic moment indicates that, in principle, there is the potentiality to measure both strength and orientation of very small magnetic dipoles using a method much faster than holography and therefore less affected by drift limitations. On the other hand it also lends itself for studies of the magnetic dipole evolution with time at very short time scales.

The same task can be performed by means of a sequence of two distinct circular-sector transformations in cascade, with factors $n_1$ and $n_2$ satisfying the condition $n_1 \cdot n_2 = -1$. As a matter of fact, the cascade of two transformation implements the total coordinate change $(\rho, \varphi) = (c \cdot r^{1/N}, \vartheta / N)$, being $N = n_1 \cdot n_2$. Even if each transformation requires two separated elements, i.e. transformer and phase-corrector, on the other hand the first phase-corrector and the second transformer can be integrated into a single element, therefore reducing to three the total number of phase plates to be realized. In this case, it is worth noting the change of sign in the radial exponent with respect to the single transformation, signifying that the composition of two Hossack's (anti-holomorphic) transformations is not a Hossack's transformation as well, as it is a holomorphic mapping (see Supplementary Information S5).

To conclude, we have demonstrated the deep link between Hossack's transformations and harmonic phases landscape for electrons, paving the way to the use of simple electrodes to produce an almost arbitrary conformal mapping. In electron microscopy this could be used as a scheme for electron beam shaping or for the optimal measurement of quantum quantities.

In particular we singled out the conformal mappings that permit to explicitly transform multipolar phases into a spot with position dependent on the multipole strength and orientation, demonstrating the possibility to measure astigmatism and dipole moment in a direct and compact manner. This in turn paves the way to spatially "compressed" measurement of magnetic dipoles of small magnetic objects.

**ACKNOWNLEDGEMENTS**

V.G. would like to thank Prof. Giulio Pozzi for fruitful discussions. This work is supported by Q-SORT, a project funded by the European Union's Horizon 2020 Research and Innovation Program under grant agreement No. 766970.

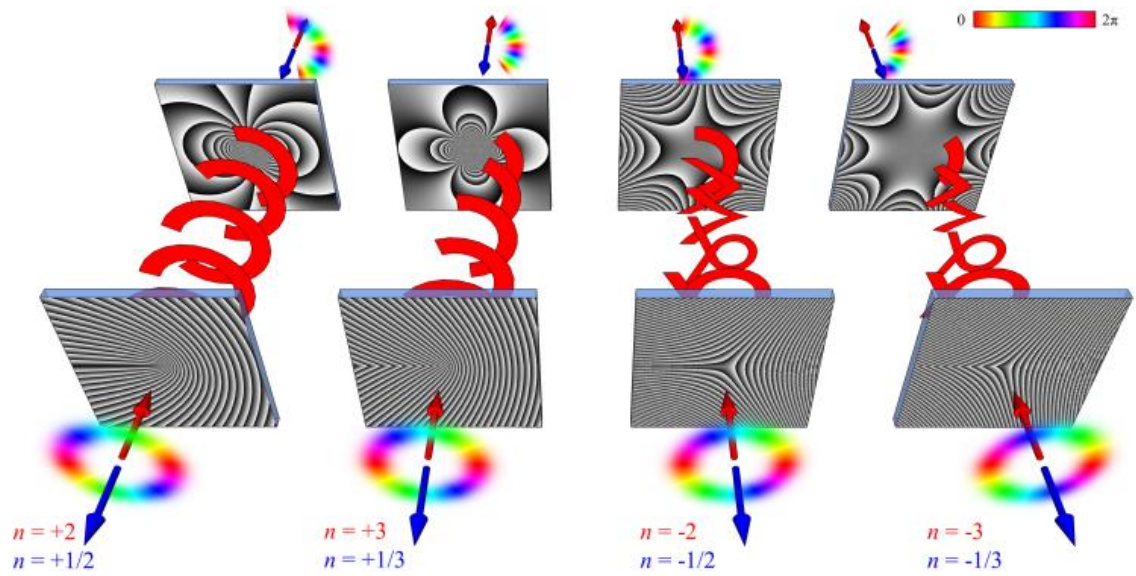

**Figure 1**. Schemes of *n*-fold circular-sector transformation of a vortex wavefunction carrying OAM $\ell=+2$, for $n=\{\pm 2, \pm 3\}$ (red arrows direction). The same configurations perform the inverse transformations, i.e. $n=\{\pm 1/2, \pm 1/3\}$, when exploited in reverse (blue arrows direction).

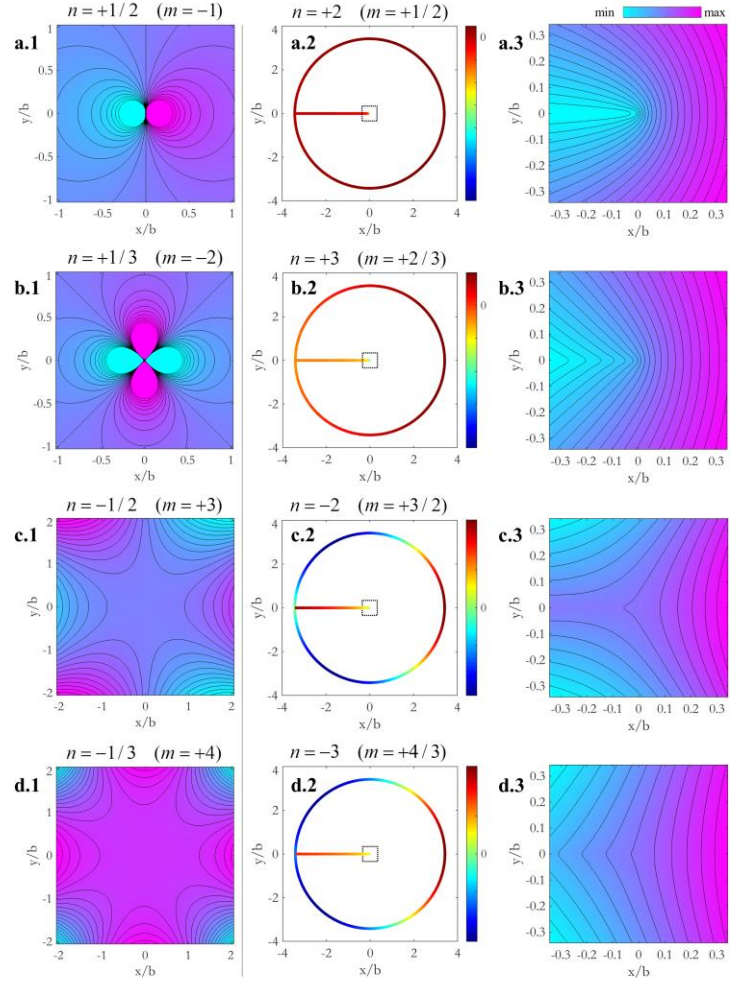

**Figure 2**. Projected charges and potentials implementing the phase elements for the circular-sector transformations (CST) depicted in Fig. 1. CST with $n=2$ (a) and $n=+3$ (b). The phases of the second elements (phase-correctors – implementing CST with $n=+1/2$ and $n=+1/3$) are proportional to the projected potentials generated by an electric dipole (a.1) and an electric quadrupole (b.1), respectively. For the first elements (transformers), the total charge distributions (a.2, b.2) are made of a linear charge density $\rho_L \propto r^{-1/n}$ along the negative x-axis, and of a ring charge distribution $\rho_R \propto \cos(m\vartheta)$, $m=1-1/n$, generating the projected potentials of fractional multipoles (a.3, b.3). CST with $n=-2$ (c) and $n=-3$ (d). In this case, the phase-correctors (implementing CST with $n=-1/2$ and $n=-1/3$) resemble the projected potentials of a magnetic sextupole (c.1) and of a magnetic octupole (d.1), respectively. As in the previous case, the projected potentials (c.3, d.3) of the transformers are generated by the combination of linear and ring charge distributions (c.2, d.2). All the projected potentials in the third column refer to the black dashed boxes in the projected charge graphs.

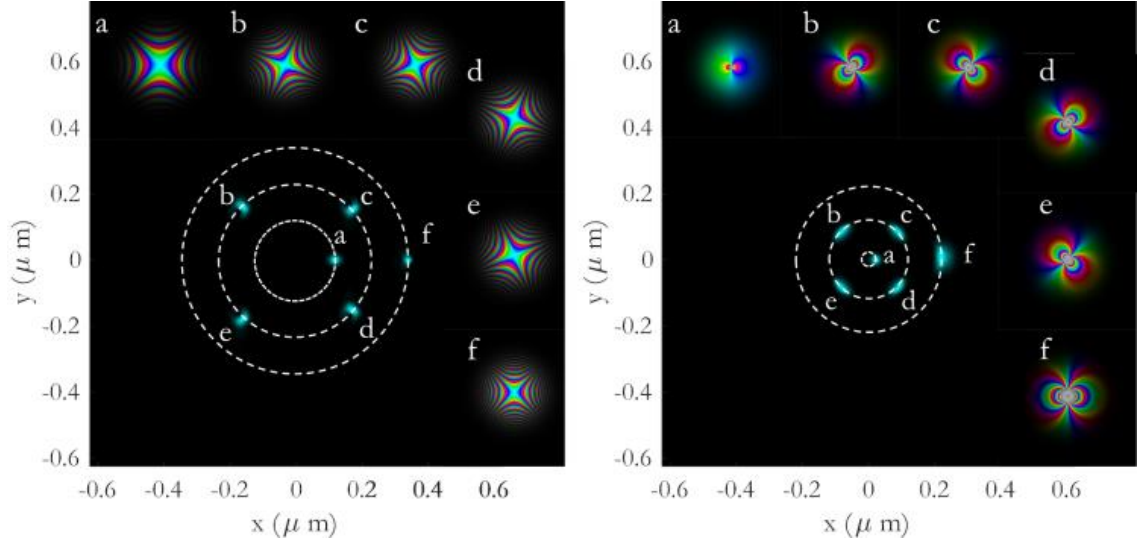

**Figure 3**. (On the left) Numerical simulations of astigmatism measurement using a circular-sector transformation with factor *n*=-1/2 and parameters *a*=4.0 µm, *b*=2.5 µm, *f*=50 cm, $\lambda$=1.9687 pm (300 keV). The input phase is transformed into a far-field spot with position depending on the astigmatism value (30 nm (a), 60 nm (b-e), 90 nm (f) for a focal length of 3.33 mm) and in-plane rotation (b-e). (On the right) Dipole sorting using a circular sector transformation with factor *n*=+1, and same design parameters. The input dipole phase is transformed into a far-field spot with position depending on the dipole moment value ($1 \cdot 10^6 \mu_B$ (a), $5 \cdot 10^6 \mu_B$ (b-e), $10 \cdot 10^6 \mu_B$ (f)) and in-plane rotation (b-e). In each subfigure, the inset plots show the input fields, brightness and colours referring to intensity and phase, respectively.

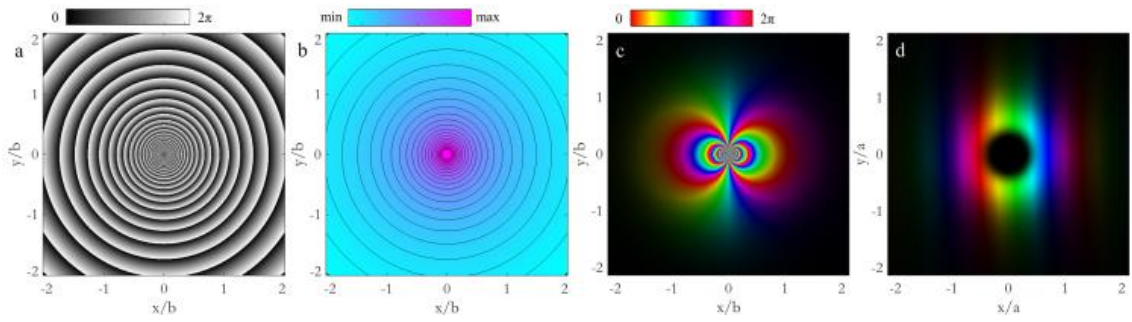

**Figure 4.** Conformal transformation of a dipole field into a linear phase contribution using a circular-sector transformation with factor *n*=+1. Comparison between the transformer phase element (a) and the projected potential (b) of a monopole placed at the origin. Input dipole field (c) and output transformed field (d). Brightness and colours refer to intensity and phase, respectively.

# SUPPLEMENTARY INFORMATION

**S1. Circular-sector transformations as solutions of Laplace's equation in 2D**

In this section, we demonstrate how the general solution of 2D Laplace equation in polar coordinates provides a potential profile inducing a circular-sector transformation. As a matter of fact, a general solution of Laplace's equation $\nabla^2 \Omega = 0$ in polar coordinates $(r, \vartheta)$ is given by:

$$\Omega(r,\vartheta) = A \cdot r^m \cdot \cos(m\vartheta + \vartheta_0) \tag{1}$$

The solution of Fresnel-Kirchhoff integral in the stationary phase approximation expresses a mapping between the wavefunction on the input plane $(r, \vartheta)$ and the transformed wavefunction on the output plane $(\rho, \varphi)$ upon propagation after a phase element described by eq. (1) [1]. This mapping is related to the gradient of the phase $\Omega$ as it follows:

$$\overline{\rho} = \frac{f}{k} \nabla \Omega \tag{2}$$

where $\overline{\rho} = \rho(\cos\varphi, \sin\varphi)$. The last relation allows obtaining the conformal transformation achieved by the given phase element, i.e. the functional relation between the two polar reference frames $(r, \vartheta)$ and $(\rho, \varphi)$ on the input and output planes, respectively. From eq. (1) it follows:

$$\begin{cases} \dfrac{\partial \Omega}{\partial x} = Amr^{m-1} \cos(\vartheta - m\vartheta - \vartheta_0) \\ \dfrac{\partial \Omega}{\partial y} = Amr^{m-1} \sin(\vartheta - m\vartheta - \vartheta_0) \end{cases} \tag{3}$$

After inserting the previous derivatives in eq. (2) it is straightforward to obtain:

$$(\rho, \varphi) = \left( \frac{f}{k} Amr^{m-1}, (1-m)\vartheta - \vartheta_0 \right) \tag{4}$$

Under the substitution $m = 1 - 1/n$, eqs. (4) describe a conformal circular-sector transformation by a factor $n$, mapping the whole circle onto a $2\pi/n$ circular sector [2]:

$$(\rho, \varphi) = \left( a(r/b)^{-1/n}, \vartheta/n - \vartheta_0 \right) \tag{5}$$

being $(a, b)$ scaling parameters. With the previous definitions, eq. (1) can be rewritten in the form:

$$\Omega(r, \vartheta) = \frac{kab}{f} \left( \frac{r}{b} \right)^{1 - \frac{1}{n}} \frac{\cos\left( \left(1 - \frac{1}{n}\right) \vartheta - \vartheta_0 \right)}{1 - \frac{1}{n}} \tag{6}$$

The corresponding phase-corrector is the optical element performing the inverse optical transformation, i.e. mapping a circular sector with angular amplitude $2\pi/n$ onto the whole circle. Its phase function is provided by a circular-sector transformer (eq. (6)) under the substitution $n \to 1/n$, $a \leftrightarrow b$, that is:

$$\Omega(\rho, \varphi) = \frac{kab}{f} \left( \frac{\rho}{a} \right)^{1-n} \frac{\cos\left( (1-n)\varphi + n\vartheta_0 \right)}{1 - n} \tag{7}$$

Analogously, it can be demonstrated that the trivial circular-sector transformation given by $n=+1$ corresponds to the logarithmic radial solution of 2D Laplace's equation, that is:

$$\Omega(r, \vartheta) = A \cdot \log\left( \frac{r}{b} \right) \tag{8}$$

As a matter of fact, the partial derivatives are given by:

$$\begin{cases} \dfrac{\partial \Omega}{\partial x} = \dfrac{A}{r} \cos(\vartheta) \\ \dfrac{\partial \Omega}{\partial y} = \dfrac{A}{r} \sin(\vartheta) \end{cases} \tag{9}$$

As done above, after inserting the previous expressions into eq. (2) it is straightforward to obtain the relations between the two reference frames:

$$(\rho, \varphi) = \left( \frac{f}{k} \frac{A}{r}, \vartheta \right) \qquad (10)$$

which describes a conformal circular-sector transformation (eq. (5)) with $n=+1$ (i.e. $m=0$) and the definition $A = kab/f$.

**S2. Projected potential produced by a multipole field**

In this section we demonstrate that the projected potential generated by an electric multipole of order $2^l$ exhibits the same spatial dependence as in eq. (1), with negative $m=-l$, therefore the transferred phase term induces a $n$-fold circular-sector transformation, with $n = 1/(1-m)$, as shown above.

The Coulomb potential of a continuous charge distribution $\rho(\bar{s})$ is given by:

$$\varphi(\bar{R}) = \frac{1}{4\pi\varepsilon_0} \int dV \frac{\rho(\bar{s})}{|\bar{R} - \bar{s}|} \qquad (11)$$

Considering all the charges are limited to some compact volume, while we want to know the potential far away from that volume, then we may expand the denominator in the Coulomb potential according to [3]:

$$\frac{1}{|\bar{R} - \bar{s}|} = \sum_{l=0}^{+\infty} \frac{s^l}{R^{l+1}} P_l(\cos \gamma) \qquad (12)$$

where $\gamma$ is the angle formed by the vectors $\bar{R}$ and $\bar{s}$. Then the Coulomb potential can be expressed as:

$$\varphi(\overline{R}) = \sum_{l=0}^{+\infty} \frac{1}{4\pi\varepsilon_0 R^{l+1}} \int dV \rho(\overline{s}) s^l P_l(\cos\gamma) \tag{13}$$

Consequently, we can decompose the potential into a series of multipole potentials:

$$\varphi(\overline{R}) = \sum_{l=0}^{+\infty} \frac{M_l(\overline{R})}{4\pi\varepsilon_0 R^{l+1}} \tag{14}$$

where $M_l(\overline{R})$ is the $2^l$-pole moment of the charge distribution in the direction of the vector $\overline{R}$. It is useful to consider the expansion of the Legendre polynomial $P_l(\cos\gamma)$ into spherical harmonics $Y_l^m(\vartheta,\varphi) = P_l(\cos\vartheta)e^{im\varphi}$, that is:

$$P_l(\cos\gamma) = \frac{4\pi}{2l+1} \sum_{m=-l}^{+l} Y_l^m(\vartheta,\varphi) Y_l^m(\theta,\phi)^* \tag{15}$$

where $(\vartheta,\varphi)$ and $(\theta,\phi)$ are the angles of vectors $\overline{R}$ and $\overline{s}$ in spherical coordinates, respectively. Substituting the last expansion in eq. (13), we obtain the following expression for eq. (14):

$$\varphi(\overline{R}) = \frac{1}{4\pi\varepsilon_0} \sum_{l=0}^{+\infty} \sum_{m=-l}^{+l} M_l^m \frac{Y_l^m(\vartheta,\varphi)}{R^{l+1}} \tag{16}$$

where the spherical harmonics of multipoles are given by:

$$M_l^m = \frac{4\pi}{2l+1} \int dV s^l Y_l^m(\theta,\phi) \rho(\overline{s}) \tag{17}$$

If we limit our analysis to an in-plane multipole, then the corresponding $2^l$-pole contribution can be expressed as:

$$\varphi_{2l}(R,\vartheta,\varphi) = \frac{1}{4\pi\varepsilon_0} M_l^l \frac{P_l(\cos\vartheta)}{R^{l+1}} \cos(l\varphi) \tag{18}$$

The corresponding projected potential, as a result of the integration in $z = R\cos\vartheta$, in given by:

$$\varphi_{2l}^z(\vartheta,\varphi) = \frac{1}{4\pi\varepsilon_0}\int_{-\infty}^{+\infty}dz M_l^l \frac{P_l(\cos\vartheta)}{R^{l+1}}\cos(l\varphi) = -\frac{1}{4\pi\varepsilon_0}\cos(l\varphi)\int_0^{\pi}d\vartheta M_l^l \frac{P_l(\cos\vartheta)}{R^l}\sin\vartheta$$
$$= -\frac{1}{4\pi\varepsilon_0}\frac{\cos(l\varphi)}{\rho^l}\int_0^{\pi}d\vartheta M_l^l \frac{P_l(\cos\vartheta)}{(1+\cos^2\vartheta)^{1/2}}\sin\vartheta = \frac{1}{4\pi\varepsilon_0}\frac{\cos(l\varphi)}{\rho^l}C_l^l \quad (19)$$

where we introduced the polar coordinates $(\rho,\varphi)$ on the plane $z=0$.

**S3. Charge distribution calculation**

In this section, we perform the analytical calculation of the charge distribution originating the desired circular-sector transformation when $m>1$, i.e. $n<0$, and in the case $0<m<1$, i.e. integer positive $n$. For fractional positive $n$, i.e. integer $m \leq 0$, the phase is proportional to the projected potential of an electric multipole generated by the proper distribution of charges around the origin, i.e. electric monopole ($n=+1$, $m=0$) electric dipole ($n=+1/2$, $m=-1$), electric quadrupole ($n=+1/3$, $m=-2$), etc.), as shown in the previous section.

The calculation of the charge density is based on the relationship between the two-dimensional phase shift and the projected charge. A quick approach is to take the Laplacian of the phase element, which corresponds to the projected charge density:

$$\nabla^2\phi(r,\vartheta) = -\frac{\rho(r,\vartheta)}{\varepsilon_0} \quad (20)$$

In polar coordinates we have:

$$\left[\frac{1}{r}\frac{\partial}{\partial r}\left(r\frac{\partial}{\partial r}\right) + \frac{1}{r^2}\frac{\partial^2}{\partial\vartheta^2}\right]\phi(r,\vartheta) = -\frac{\rho(r,\vartheta)}{\varepsilon_0} \quad (21)$$

We insert in the previous equation the phase function:

$$\phi(r,\vartheta) = Ar^m\cos(m\vartheta)\Theta(R-r), \quad \vartheta\in(-\pi,\pi] \quad (22)$$

where $\Theta(\cdot)$ is the Heaviside function: $\Theta(x)=1$ when $x \geq 0$, $\Theta(x)=0$ otherwise. When $m$ is fractional, a discontinuity line appears at $\vartheta = \pi$ for the first derivative in the azimuthal direction. The two contributions on the first side of eq. (21) can be easily calculated as it follows:

$$\frac{1}{r}\frac{\partial}{\partial r}\left(r\frac{\partial}{\partial r}\right)\phi(r,\vartheta) = Am^2 r^{m-2}\cos(m\vartheta)\Theta(R-r) - 2A(m+1)R^{m-1}\cos(m\vartheta)\delta(R-r) \quad (23)$$

$$\frac{1}{r^2}\frac{\partial^2}{\partial \vartheta^2}\phi(r,\vartheta) = -Am^2 r^{m-2}\cos(m\vartheta)\Theta(R-r) + 2Amr^{m-2}\sin(m\pi)\delta(\vartheta-\pi)\Theta(R-r) \quad (24)$$

where $\delta(\cdot)$ is the Dirac delta function. Therefore the total projected charge density $\rho$ can be split into two contributions, in the specific a ring charge distribution $\rho_R$ and a line charge distribution $\rho_L$, given by:

$$\rho_R(\vartheta) = Q_R \cos(m\vartheta)\delta(R-r) \quad (25)$$

$$\rho_L(r) = Q_L \cdot r^{m-1}\Theta(R-r)\delta(\vartheta-\pi) \quad (26)$$

where $Q_R = 2\varepsilon_0 A(m+1)R^{m-1}$, $Q_L = -2\varepsilon_0 Am\sin(m\pi)$. The radius $R$ of the charged ring can be selected in order to satisfy the total charge neutrality. For integer positive $m$, the phase resembles the one of an external magnetic multipole, then only the ring distribution is required, since there is no discontinuity in the azimuthal direction. On the other hand, for negative integer $m$, the phase is generated by a proper distribution of electric poles around the origin, as mentioned above.

For a given charge density, the generated potential has been calculated applying the integral:

$$\phi(\bar{r},z) = \frac{1}{4\pi\varepsilon_0}\iint \frac{\rho(\bar{r}')}{\sqrt{|\bar{r}-\bar{r}'|^2+z^2}} r'dr'd\vartheta' \quad (27)$$

Finally, the projected potential is given by [4]:

$$\phi^z(r,\vartheta) = \int_{-\infty}^{+\infty} \phi(\bar{r},z)dz \tag{28}$$

## S4. Sorting of multipole phase profiles with a circular-sector transformation

We consider a wavefunction with the input phase:

$$\Phi^{in}(r,\vartheta) = Ar^m \cos(m(\vartheta-\vartheta_0)) \tag{29}$$

After applying a circular-sector transformation by a factor $n$, and parameters $(a, b)$, that is:

$$(\rho,\varphi) = \left(a\left(\frac{r}{b}\right)^{-\frac{1}{n}}, \frac{\vartheta}{n}\right) \tag{30}$$

then the output phase is given by:

$$\Phi^{out}(\rho,\varphi) = A^{out}\rho^{-mn}\cos(mn\varphi - m\vartheta_0) \tag{31}$$

where $A^{out} = Ab^m/a^{-mn}$. With the choice $n = -1/m$, eq. (31) reduces to the plane wave term:

$$\Phi^{out}(\rho,\varphi) = A^{out}\rho\cos(\varphi + m\vartheta_0) \tag{32}$$

The far-field at a distance $z$ exhibits a spot forming an angle $-m\vartheta_0$ with the x-axis and shifted of $zA^{out}/k$ with respect to the optical axis. In the main text, we considered the following examples:

- Input external quadrupole field (astigmatism): $m=+2$. Sorting transformation: $n=-1/2$. The far-field spot is tilted of an angle $\theta(A) = Ab^2/(ka)$ with respect to the optical axis, and forms an angle with the x-axis equal to $-2\vartheta_0$.

- Input internal dipole: $m=-1$. Sorting transformation: $n=+1$. The far-field spot is tilted of an angle $\theta(A) = Ab^{-1}/(ka)$, and forms an angle with the x-axis equal to $\vartheta_0$.

**S5. Sorting of multipole phase profiles with a cascade of two circular-sector transformations**

We consider the initial phase profile:

$$\Phi^{in}(r,\vartheta) = Ar^m \cos(m(\vartheta - \vartheta_0)) \qquad (33)$$

If we apply a circular-sector transformation by a factor $n_1$, and parameters $(a_1, b_1)$:

$$(\rho', \varphi') = \left( a_1 \left(\frac{r}{b_1}\right)^{-\frac{1}{n_1}}, \frac{\vartheta}{n_1} \right) \qquad (34)$$

followed by a second circular-sector transformation by a factor $n_2$, and parameters $(a_2, b_2)$:

$$(\rho, \varphi) = \left( a_2 \left(\frac{\rho'}{b_2}\right)^{-\frac{1}{n_2}}, \frac{\varphi'}{n_2} \right) \qquad (35)$$

we obtain the new kind of transformation:

$$(\rho, \varphi) = \left( \frac{a_2 a_1^{-1/n_2}}{b_2^{-1/n_2}} \left(\frac{r}{b_1}\right)^{\frac{1}{n_1 n_2}}, \frac{\vartheta}{n_1 n_2} \right) \qquad (36)$$

where it must be noticed the change of sign in the exponent with respect to each separate transformation. This phase change makes this mapping structurally (holomorphic) different from each of the separate mappings (anti-holomorphic).

The final phase term is given by:

$$\Phi^{out}(\rho, \varphi) = A^{out} \rho^{mn_1 n_2} \cos(mn_1 n_2 \varphi - m\vartheta_0) \qquad (37)$$

where $A^{out} = Ab_1^m b_2^{-mn_1} / (a_1^{-mn_1} a_2^{mn_1 n_2})$. With the choice $n_1 n_2 = +1/m$, and considering an input dipole phase function, i.e. $m=-1$, then eq. (32) reduces to the plane wave term

$$\Phi^{out}(\rho,\varphi) = A^{out}\rho\cos(\varphi+\vartheta_0) \tag{38}$$

where $A^{out} = Ab_1^{-1}b_2^{n_1}/(a_1^{n_1}a_2)$.

In Fig. S1, the transformation of an input dipole phase profile into an output spot is shown, by applying in sequence a set of transformations with factors $(n_1, n_2)=(-2, +1/2)$.

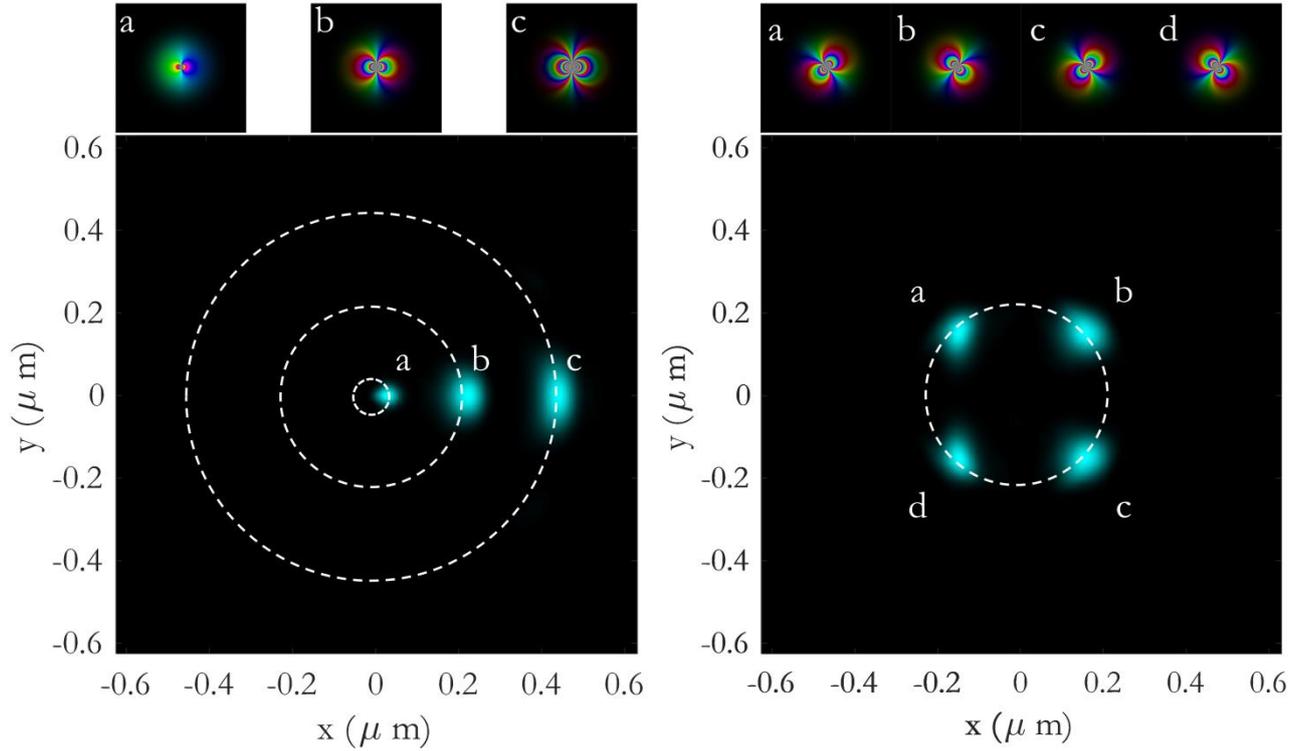

**Figure S1**. Numerical simulations of dipole sorting using a sequence of two circular-sector transformations with parameters $(n_1, n_2)=(-2, +1/2)$. The input dipole phase is transformed into a far-field spot with position depending on the dipole strength (left panel, $1\cdot 10^6$ μ$_B$ (a), $5\cdot 10^6$ μ$_B$ (b), $10\cdot 10^6$ μ$_B$ (c)) and rotation (right panel, $5\cdot 10^6$ μ$_B$). In each subfigure, the inset plots show the input dipole field, where brightness and colours refer to intensity and phase, respectively.